\documentclass[showpacs,aps,twocolumn]{revtex4}
\usepackage{epsfig}
\usepackage{graphicx}
\usepackage{amsmath,amssymb,amsfonts}
\usepackage{array}
\usepackage{url}
\usepackage{hyperref}
\usepackage{multirow}
\usepackage{float}
\usepackage{lineno}
\usepackage{xspace}
\usepackage[usenames,dvipsnames]{color}

\newcommand{\bef}{\begin{figure}}
\newcommand{\eef}{\end{figure}}
\newcommand{\bc}{\begin{center}}
\newcommand{\ec}{\end{center}}

\newcommand{\be}{\begin{equation}}
\newcommand{\ee}{\end{equation}}
\newcommand{\bea}{\begin{eqnarray}}
\newcommand{\eea}{\end{eqnarray}}

\def\ba{\begin{eqnarray}}
\def\ea{\end{eqnarray}}
\begin{document}

\title{Hadron gas in the presence of a magnetic field using non-extensive statistics: A transition from diamagnetic to paramagnetic system}
\author{Girija Sankar Pradhan$^{1}$}
\author{Dushmanta Sahu$^{1}$}
\author{Suman Deb$^{1}$}
\author{Raghunath Sahoo$^{1,2,}$\footnote{Corresponding Author Email: Raghunath.Sahoo@cern.ch}\footnote{We would like to dedicate this paper to the loving memory of Prof. Jean Cleymans, who introduced us non-extensivity and 
we had the fortune of working with him on several publications. }
}
\affiliation{Department of Physics, Indian Institute of Technology Indore, Simrol, Indore 453552, India}
\affiliation{$^{2}$CERN, CH 1211, Geneva 23, Switzerland}

\begin{abstract}
Non-central heavy-ion collisions at ultra-relativistic energies are unique in producing magnetic fields of the largest strength in the laboratory. Such fields being produced at the early stages of the collision could affect the properties of Quantum Chromodynamics (QCD) matter formed in the relativistic heavy-ion collisions. The transient magnetic field leaves its reminiscence, which in principle, can affect the thermodynamic and transport properties of the final state dynamics of the system. In this work, we study the thermodynamic properties of a hadron gas in the presence of an external static magnetic field using a thermodynamically consistent non-extensive Tsallis distribution function. Various thermodynamical observables such as energy density ($\epsilon$), entropy density ($s$), pressure ($P$) and speed of sound ($c_{\rm s}$) are studied. Investigation of magnetization ($M$) is also performed and this analysis reveals an interplay of diamagnetic and paramagnetic nature of the system in the presence of a magnetic field of varying strength. Further, to understand the system dynamics under equilibrium and non-equilibrium conditions, the effect of the non-extensive parameter ($q$) on the above observables is also studied. 

\end{abstract}
\date{\today}
\maketitle{}

\section{Introduction}
\label{intro}

One of the fundamental laws of electromagnetism is the Biot-Savart law, which states that the electric charge in motion can produce a magnetic field ($B$). In the same analogue, magnetic fields of larger strengths are created in the relativistic colliders by the spectator protons in peripheral heavy-ion collisions. Such colliders provide us a unique tool to explore and understand the Quantum Chromodynamics (QCD) phase diagram for a wide range of temperature ($T$) and baryon-chemical potential ($\mu_{B}$). $T$ and $\mu_{B}$ serve as vital parameters in understanding the equation of state (EoS) governing the evolution of the system formed in such collider experiments. Additionally, the fast and oppositely directed motion of the non-central relativistic heavy-ion collisions (HIC) can create an extremely strong transient electromagnetic field due to the relativistic motion of the colliding beams carrying a large positive electric charge (or spectator protons) ~\cite{Warringa:2012bq,Fukushima:2010vw,Kharzeev:2007jp}. In experiments such as STAR at Relativistic Heavy Ion Collider (RHIC) and ALICE at the Large Hadron Collider (LHC), where heavy-ions are collided to study the resulting QCD matter, the maximum magnetic field generated can be of the order; $eB \thicksim(m_{\pi}^{2}) \thicksim 10^{18}$G ~\cite{Bzdak:2011yy} and $eB \thicksim(15m_{\pi}^{2})$ respectively, where $m_{\pi}$ is the mass of a pion ~\cite{Skokov:2009qp,Deng:2012pc}. This is astronomically larger than the strongest man-made steady magnetic field in the laboratory. Strong magnetic fields are also expected to be present in dense neutron stars ~\cite{Duncan:1992hi,Dey:2002pc} and might have been present during the electroweak transition in the early universe ~\cite{Vachaspati:1991nm,Bhatt:2015ewa}. The magnetic field created on earth in the form of an electromagnetic shock wave is $\thicksim10^{7}$ G and the estimated value of the magnetic field inside a neutron star is $10^{10} -10^{13}$ G ~\cite{Konar:2017kty,Reisenegger:2013faa}. In comparison to these values, the electromagnetic field created in a heavy-ion collision is perhaps the strongest magnetic field that has ever been created in nature. This makes it important to study the effect of the magnetic field on the hot and dense matter formed in such collisions. Thus, besides T and $\mu_{B}$, the parameter $B$ can affect the EoS and hence it can have a vital role to play in our understanding of the phase diagram. 

Although the strength of the electromagnetic field created in the HIC is very large, it exists for a very short duration. For instance, the maximum strength of electromagnetic field created for Au + Au at $\sqrt {s_{\rm NN}} =$ 200 GeV is of the order of $ 5m_{\pi}^{2} $ but the duration it lasts for is only about 0.3 fm/c ~\cite{Kharzeev:2007jp,Skokov:2009qp}. Thus, the strength of the electromagnetic field created in the initial off-central collisions almost decays with the evolution of the matter and practically, the initially produced magnetic field is not observed in the final state. However, as mentioned earlier, the strength of the initially produced magnetic field is so large that it can affect the EOS of the matter produced in the collisions and whose effect should be observed in the final state observables. Furthermore, the existence of electrical conductivity will delay the decay of this transient magnetic field significantly, which might mean that even in the hadronic phase, a relatively small magnetic field can be present. Present work is an attempt to explore this effect of initial magnetic field on the final state observables in an away from equilibrium regime, as one considers non-central heavy-ion collisions, where the particle transverse momentum spectra are better described by non-extensive Tsallis statistical distribution function.

In literature, there are several studies on the effect of external magnetic fields on the hadron gas ~\cite{Kadam:2019rzo,Endrodi:2013cs, Tawfik:2016lih}. However, as mentioned above, the present work deals with the study of the hadron gas system away from equilibrium and under external magnetic fields. As we know, it is observed that at RHIC ~\cite{ Abelev:2006cs,Adare:2011vy} and the LHC ~\cite{Aamodt:2011zj,Abelev:2012cn,Abelev:2012jp,Chatrchyan:2012qb,Bhattacharyya:2015hya} energies, the transverse momentum spectra of the produced particles in $pp$ and peripheral heavy-ion collisions deviate from the standard Boltzmann-Gibbs (BG) distribution. The BG distribution describes the low-$p_{\rm T}$ part of the transverse momenta ($p_{\rm T}$) spectra very well, whereas a power-law type distribution explains the high-$p_{\rm T}$ part of the transverse momentum spectra. Traditionally, two different types of distribution functions were used to describe the entire range of $p_{\rm T}$-spectra. However, it has been seen that the Tsallis non-extensive distribution describes $p_{\rm T}$-spectra very well for the whole $p_{\rm T}$ range. This has motivated us to use a thermodynamically consistent form of the Tsallis distribution function for the present work. Such a distribution function has been used extensively to study the thermodynamical properties of the systems formed in heavy-ion collisions ~\cite{Deb:2019yjo,Deb:2021gzg,Sahu:2020swd}. The Tsallis distribution function introduces a fitting parameter called the non-extensive parameter ($q$), which gives the degree of deviation from equilibrium. $q = 1$ suggests the equilibrium condition (BG scenario) and it can have values between the range of 1 to 11/9 ~\cite{Beck:2009uy} depending upon the degree of deviation of the system. Generally, the systems formed in $pp$ and peripheral heavy-ion collisions are away from equilibrium. Thus, it would be appropriate to take the non-extensive parameter into consideration for the current study.

As the magnetic field is transient in nature, although of very high magnitude, in a given theoretical model, it is very difficult 
to incorporate it at the initial pre-equilibrium phase and evolve through various complex processes. For the sake of simplicity,
in the present work, we take a static magnetic field and instead of a complex dynamical QCD system, we consider a final
state hadronic system to study the possible effects. To do this, analogous to an away from equilibrium system, like 
the one usually achieved in non-central heavy-ion collisions, we formulate the problem with a non-extensive statistical distribution function with a magnetic field and study various thermodynamic properties.

The present paper illustrates the thermodynamic observables, including energy density, pressure, entropy density, magnetization and the speed of sound in a hadron gas by using the framework of Tsallis non-extensive statistics in a finite and static magnetic field. The section \ref{formulation} briefly describes the formulation for including the magnetic field and estimating different thermodynamical observables in Tsallis non-extensive formalism. Section \ref{vacpressure} explicitly describes the regularization of vacuum pressure. We report the results and discussion in section \ref{res}, and finally, in section \ref{sum}, we summarize our findings.  Further, in the appendix, we show the thermodynamic consistency of the Tsallis distribution function in the presence of an external magnetic field.

\section{Formulation}
\label{formulation}

The dispersion relation of a particle moving in a nonzero magnetic field is given by ~\cite{Kadam:2019rzo,Endrodi:2013cs,Tawfik:2016lih,Landau,Bali:2014kia,Fraga:2008qn},

\begin{equation}
\label{eq1}
E = \bigg[ p_{z}^{2}+m^{2}+2\vert e \vert (k - s_{z} + \frac{1}{2})B \bigg]^{1/2},
\end{equation}   

where $k$ is any positive integer corresponding to allowed Landau levels, $e$ is the charge of the particle under consideration, $s_{z}$ is the component of spin along the z-direction and $B$ is the magnetic field strength. Here $p_{z}$ is the momentum of the particle along the z-direction and $m$ is its corresponding mass.

Thus, the energy for neutral ($n$) and charged particles ($c$) in the presence of an external magnetic field are respectively given by,

\begin{equation}
\label{eq2}
E_{i,n}= \sqrt{p^{2}+m_{i}^{2}}
\end{equation}                            
                        
\begin{equation}
\label{eq3}
E_{i,c}(p_{z},k,s_{z}) = \sqrt{p_{z}^{2}+m_{i}^{2}+2|e_{i}|B(k+1/2-s_{z})}.
\end{equation}                            
                         
In the present study, going inline with Ref ~\cite{Endrodi:2013cs}, $i$ in both Eq.~\ref{eq2} and ~\ref{eq3} refers to the hadrons upto the mass of $1.2$ GeV.
 
Further, we introduce the thermodynamically consistent Tsallis distribution function as ~\cite{Tsallis:1987eu,Tsallis:1998ws,Cleymans:2011in,Azmi:2015xqa},

 \begin{equation}
 \label{eq4}
 f(E,q,T,\mu) = \frac {1}{exp_{q}(\frac{E - \mu}{T})},
 \end{equation} 
  
with, 
\begin{equation}
\label{eq5}
\exp_{q}(x) \equiv
 \begin{cases}
  [1+(q-1)x]^{\frac{1}{q-1}}    & \quad \text{if } x > 0\\
  [1+(1-q)x]^{\frac{1}{1-q}}   & \quad \text{if } x \le 0\\
\end{cases}
\end{equation}

where, $x = (E-\mu)/T$. Here, $E$, $\mu$, $T$ and $q$ are the energy, chemical potential, temperature and the non-extensive parameter respectively. It is worth noting that in the limit, $q \rightarrow$ 1, Eq. \eqref{eq5} reduces to the standard exponential function,

\begin{eqnarray*}
\lim_{q \to 1} \exp_q(x) \rightarrow \exp(x).
\end{eqnarray*}

The thermodynamical quantities such as number density, energy density, pressure, and entropy density in non-extensive statistics are given by ~\cite{Cleymans:2011in,Azmi:2015xqa,Cleymans:2012ya},
  
\begin{equation}
\label{eq6}
n = g\int \frac{d^{3}p}{(2\pi)^{3}}[1+(q-1)\frac{E-\mu}{T}]^{\frac{-q}{q-1}} 
\end{equation}  

\begin{equation}
\label{eq7}
\epsilon = g\int \frac{d^{3}p}{(2\pi)^{3}}E[1+(q-1)\frac{E-\mu}{T}]^{\frac{-q}{q-1}} 
\end{equation}  

\begin{equation}
\label{eq8}
P = g\int \frac{d^{3}p}{(2\pi)^{3}}\frac{p^{2}}{3E}[1+(q-1)\frac{E-\mu}{T}]^{\frac{-q}{q-1}} 
\end{equation}  

\begin{equation}
\label{eq9}
s = -g\int \frac{d^{3}p}{(2\pi)^{3}} \bigg[\frac{f- f^{q}}{1-q} -f\bigg],
\end{equation}

where $n$, $\epsilon$, $P$, $s$ and $g$ are the number density, energy density, pressure, entropy density, and degeneracy, respectively. In presence of a finite magnetic field, the phase-space integral in the above equations are expressed as a one-dimensional integral, $\frac{1}{({2\pi})^{2}}\iint dp_{x}dp_{y} = \frac{|e_{i}|B}{2\pi}$ ~\cite{Fraga:2008qn,Chakrabarty:1996te}.

Now, we can rewrite Eq.~\ref{eq7} and \ref{eq8} for charged and neutral particles in the presence of a strong magnetic field for chemical potential, $\mu = 0$ as follows,
 
\begin{equation}
\label{eq10}
\epsilon_{c} = \sum_{i}\sum_{k}\sum_{s_{z}}\frac{g_{i}|e_{i}|B}{(2\pi)^{2}} \int dp_{z}E_{i,c}\bigg[1+(q-1)\frac{E_{i,c}}{T}\bigg]^{\frac{-q}{q-1}}
\end{equation} 

\begin{equation}
\label{eq11}
P_{c} = \sum_{i}\sum_{k}\sum_{s_{z}}\frac{g_{i}|e_{i}|B}{(2\pi)^{2}} \int dp_{z}\frac{p^{2}}{3E_{i,c}}\bigg[1+(q-1)\frac{E_{i,c}}{T}\bigg]^{\frac{-q}{q-1}},
\end{equation}  
where, $e_{i} \neq 0$.

\begin{equation}
\label{eq12}
\epsilon_{n} = \sum_{i}g_{i}\int \frac{d^{3}p}{(2\pi)^{3}}E_{i,n}\bigg[1+(q-1)\frac{E_{i,n}}{T}\bigg]^{\frac{-q}{q-1}}
\end{equation}  

\begin{equation}
\label{eq13}
P_{n} = \sum_{i}g_{i}\int \frac{d^{3}p}{(2\pi)^{3}}\frac{p^{2}}{3E_{i,n}}\bigg[1+(q-1)\frac{E_{i,n}}{T}\bigg]^{\frac{-q}{q-1}},
\end{equation}  
where, $e_{i} = 0$.

Here, summations over $i, k, {s_{z}}$ respectively refer to all the hadrons up to the mass of $1.2$ GeV, Landau levels, and spin along the z-direction. Thus the total pressure(P) of the system is obtained by adding to the thermal part of the pressure to the vacuum part (see section\ref{vacpressure}),
\begin{equation}
\label{14}
P = P_{c}+P_{n}+\Delta P_{\text{vac}}
\end{equation}

Further, the Helmholtz free energy is written as,
\begin{equation}
\label{eq15}
f_{\rm H} = \epsilon - Ts.
\end{equation}

In the presence of constant external magnetic field B, the modified energy density and Helmholtz free energy become,
\begin{eqnarray}
\label{eq16}
\epsilon_{\text{total}} = \epsilon_{c}+\epsilon_{n}+eB(M+ \Delta M_{\text{vac}})\nonumber\\
 = \epsilon+eBM_{\text{total}},
\end{eqnarray}
where $\epsilon = \epsilon_{c}+\epsilon_{n}$ and $M_{\text{total}} = M+ \Delta M_{\text{vac}}$ and $\Delta M_{\text{vac}}$~\ref{eq47} is the vacuum pressure term due to external magnetic field present at T= 0.

\begin{equation}
\label{eq17}
f_{\rm H} =\epsilon_{\text{total}} - Ts - eBM_{\text{total}}
\end{equation}

Here, $M$ is the magnetization of the system due to external magnetic field. At the thermodynamical limit, $V\rightarrow \infty$, we can assume $f_{\rm H} = -P$ ~\cite{Endrodi:2013cs,Kadam:2019rzo}. So, from Eq.~\ref{eq17} we have,
 
\begin{equation}
\label{eq18}
eBM_{\text{total}} = \epsilon_{\text{total}} - Ts +P
\end{equation}

\begin{equation}
\label{eq19}
M_{\text{total}} =\frac{\epsilon_{\text{total}} - (Ts-P)}{eB}
\end{equation}

\begin{equation}
\label{eq20}
M_{\text{total}} =\frac{\epsilon_{\text{total}} - \epsilon}{eB},
\end{equation}
where, $\epsilon = (Ts-P)$.

In hydrodynamics, the speed of sound plays a vital role in explaining the equation of state and hence, the associated phase transition. The squared speed of sound $(c_{s}^{2})$ for a system in an external magnetic field at zero baryon chemical potential is defined as, 

\begin{equation}
\label{eq_c1}
c_{s}^{2}(T, eB) =\bigg(\frac{\partial P}{\partial \epsilon}\bigg)\bigg |_{\frac{s}{n}} = \frac{\frac{\partial P}{\partial T} + \frac{\partial P}{\partial (eB)}\frac{d(eB)}{dT}}{\frac{\partial \epsilon}{\partial T} + \frac{\partial \epsilon}{\partial (eB)}\frac{d(eB)}{dT}},
\end{equation}

where,
\begin{equation}
\label{eq_c2}
\frac{d(eB)}{dT} = \frac{s\frac{\partial n}{\partial T} - n\frac{\partial s}{\partial T}}{n\frac{\partial s}{\partial (eB)} - s\frac{\partial n}{\partial (eB)}}.
\end{equation}

 The speed of sound in the system reduces to $c_s^2 = (\frac{\partial P}{\partial \epsilon})$ for zero external magnetic field.

It is worth noting that the speed of sound can be derived from Euler's law as applied to a continuous medium without having to use thermodynamics regardless of whether the system is extensive or non-extensive ~\cite{Landaubook}.

We now proceed to calculate the renormalized vacuum pressure at a finite magnetic field in the next section.

\section{Renormalization of vacuum pressure}
\label{vacpressure}

In the previous section, we calculated the thermal part of the thermodynamic variables. In this section, we will concentrate on the vacuum contribution of these observables in the presence of an external magnetic field with the help of a regularization method. The magnetic field induces a vacuum contribution to most observables, such that, e.g., the pressure does not vanish at $T = 0$. The vacuum pressure term is divergent and it requires to be suitably regularized first. The unphysical results arise, particularly in the case of magnetic field-dependent regularization methods. Hence, it is necessary that magnetic field-dependent and independent parts must be separated distinctly through an appropriate regularization technique.

 For a charged spin-$\frac{1}{2}$ particle in the presence of an external magnetic field, the vacuum pressure is given by \cite{Kadam:2019rzo,Endrodi:2013cs,Menezes:2008qt},
 
\begin{equation}
\label{eq22}
P_{\text{vac}}(S=1/2,B)= \frac{1}{2}\sum_{k=0}^{\infty}g_{k}\frac{eB}{2\pi}\int_{-\infty}^{\infty}\frac{dp_z}{2\pi}E_{p,k}(B),
\end{equation}
where $g_{k}=2-\delta_{k0}$ is the degeneracy of $k^{\text{th}}$ Landau level. After adding and subtracting the lowest Landau level contribution term (i.e. $k=0$) from the above equation, we obtain,

\begin{equation}
\label{eq23}
P_{\text{vac}}(S=1/2,B)= \frac{1}{2}\sum_{k=0}^{\infty}2\frac{eB}{2\pi}\int_{-\infty}^{\infty}\frac{dp_z}{2\pi} \\
\bigg[E_{p,k}(B)-\frac{E_{p,0}(B)}{2}\bigg].
\end{equation}

The divergence can be regularized using dimensional regularization method~\cite{Peskin:1995}. In $d-\epsilon$ dimension Eq. ~\ref{eq23} can be written as

\begin{eqnarray}
\label{eq24}
P_{\text{vac}}(S=1/2,B)=\sum_{k=0}^{\infty}\frac{eB}{2\pi}\mu^{\epsilon}\int_{-\infty}^{\infty}\frac{d^{1-\epsilon}p_z}{(2\pi)^{1-\epsilon}} \nonumber\\ \bigg[\sqrt{p_z^2+m^2-2eBk}-\sqrt{p_z^2+m^2}\bigg],
\end{eqnarray}
 
In the above expression, $\mu$ fixes the dimension to one. The integration can be performed utilizing traditional $d-$dimensional formula~\cite{Peskin:1995,ramond}. 
\begin{equation}
\label{eq25}
\int_{-\infty}^{\infty}\frac{d^dp}{(2\pi)^d}\:\bigg[p^2+m^2\bigg]^{-A}=\frac{\Gamma[A-\frac{d}{2}]}{(4\pi)^{d/2}\Gamma[A](M^2)^{(A-\frac{d}{2})}}.
\end{equation}
 
Integration of the first term in Eq.~\ref{eq24} gives
\begin{eqnarray}
\label{eq26}
I_{1}=\sum_{k=0}^{\infty}\frac{eB}{2\pi}\mu^{\epsilon}\int_{-\infty}^{\infty}\frac{d^{1-\epsilon}p_z}{(2\pi)^{1-\epsilon}}\bigg[p_z^2+m^2-2eBk\bigg]^{\frac{1}{2}} \nonumber\\ 
=-\frac{(eB)^2}{4\pi^2}\bigg(\frac{2eB}{4\pi\mu}\bigg)^{-\frac{\epsilon}{2}}\Gamma\bigg[-1+\frac{\epsilon}{2}\bigg]\zeta\bigg[-1+\frac{\epsilon}{2},x\bigg],
\end{eqnarray}

where we denote $x\equiv\frac{m^2}{2eB}$. The Landau infinite sum has been illustrated in terms of Riemann-Hurwitz $\zeta-$function
\begin{equation}
\label{eq27}
\zeta[z,x]=\sum_{k=0}^{\infty}\frac{1}{[x+k]^z},
\end{equation}
with the expansion 

\begin{equation}
\label{eq28}
\zeta\bigg[-1+\frac{\epsilon}{2},x\bigg]\approx-\frac{1}{12}-\frac{x^2}{2}+\frac{x}{2}+\frac{\epsilon}{2}\zeta^{'}(-1,x)+\mathcal{O}(\epsilon^2)
\end{equation}
and the asymptotic behavior of the derivative~\cite{Elizalde:1985ch,dlmf},

\begin{eqnarray}
\label{eq29}
\zeta'(-1,x) &=& \frac{1}{12} -\frac{x^2}{4} + \left( \frac{1}{12} -\frac{x}{2} + \frac{x^2}{2}\right) \text{ln(x)} +
\mathcal{O}(x^{-2}).\nonumber\\ 
\end{eqnarray}

The expansion of $\Gamma$-function around some negative integers is given by
\begin{equation}
\label{eq30}
\Gamma\bigg[-1+\frac{\epsilon}{2}\bigg]=-\frac{2}{\epsilon}+\gamma-1+\mathcal{O}(\epsilon),
\end{equation}
and 
\begin{equation}
\label{eq31}
\Gamma\bigg[-2+\frac{\epsilon}{2}\bigg]=\frac{1}{\epsilon}-\frac{\gamma}{2}+\frac{3}{4}+\mathcal{O}(\epsilon),
\end{equation}

here $\gamma$ is the Euler constant.The limiting expression for natural logarithm
\begin{equation}
\label{eq32}
\lim_{\epsilon\longrightarrow 0}a^{-\epsilon/2}\approx 1-\frac{\epsilon}{2}\text{ln}(a).
\end{equation} 

Using the expansion of $\Gamma$-function and the expansion of $\zeta$-function, Eq.~\ref{eq26} can be written as
\begin{eqnarray}
\label{eq33}
I_1=-\frac{(eB)^2}{4\pi^2}\bigg[-\frac{2}{\epsilon}+\gamma-1+\text{ln}\bigg(\frac{2eB}{4\pi\mu^2}\bigg)\bigg] \nonumber\\
\bigg[-\frac{1}{12}-\frac{x^2}{2}+\frac{x}{2}+\frac{\epsilon}{2}\zeta^{'}(-1,x)+\mathcal{O}(\epsilon^2)\bigg]
\end{eqnarray}
 
The second term in Eq.~\ref{eq24} can be simplified in a similar fashion. We obtain
\begin{eqnarray}
\label{eq34}
I_{2}=\sum_{k=0}^{\infty}\frac{eB}{2\pi}\mu^{\epsilon}\int_{-\infty}^{\infty}\frac{d^{1-\epsilon}p_z}{(2\pi)^{1-\epsilon}}\bigg[p_z^2+m^2\bigg]^{\frac{1}{2}} \nonumber\\
=\frac{(eB)^2}{4\pi^2}\bigg[-\frac{x}{\epsilon}-\frac{(1-\gamma)}{2}x+\frac{x}{2}\text{ln}\bigg(\frac{2eB}{4\pi\mu^2}\bigg)+\frac{x}{2}\text{ln}(x)\bigg].\nonumber \\
\end{eqnarray}
  
Hence, the vacuum pressure in the presence of an external magnetic field becomes
\begin{eqnarray}
\label{eq35}
P_{\text{vac}}(S=1/2,B)=\frac{(eB)^2}{4\pi^2}\bigg[\zeta^{'}(-1,x)-\frac{2}{12\epsilon}-\frac{(1-\gamma)}{12}\nonumber\\
-\frac{x^2}{\epsilon}-\frac{(1-\gamma)}{2}x^2+\frac{x}{2}\text{ln}(x)\nonumber\\
+\frac{x^2}{2}\text{ln}\bigg(\frac{2eB}{4\pi\mu^2}\bigg)+\frac{1}{12}\text{ln}\bigg(\frac{2eB}{4\pi\mu^2}\bigg)\bigg].\nonumber\\
\end{eqnarray}
 
In the above expression, divergence is still present. Hence, we add and subtract $B=0$ contribution from it. In order to carry out the renormalization of the $B>0$ pressure, it is essential to determine the $B=0$ contribution. The vacuum pressure at $B=0$ in $d=3-\epsilon$ dimensions is given by 

\begin{eqnarray}
\label{eq36}
P_{\text{vac}}(S=1/2,B=0)&=& \mu^{\epsilon}\int\frac{d^{3-\epsilon}p}{(2\pi)^{3-\epsilon}}\:(p^2+m^2)^{\frac{1}{2}}\nonumber\\
&=&\frac{(eB)^2}{4\pi^2}\bigg(\frac{2eB}{4\pi\mu^2}\bigg)^{-\frac{\epsilon}{2}}\Gamma\bigg(-2+\frac{\epsilon}{2}\bigg)x^{2-\frac{\epsilon}{2}} \nonumber\\
\end{eqnarray}
 
Above equation~\ref{eq36} can be further simplified by using $\Gamma$-function expansion~\ref{eq31}
\begin{eqnarray}
\label{eq37}
P_{\text{vac}}(S=1/2,B=0)&=&-\frac{(eB)^2}{4\pi^2}x^2\bigg[\frac{1}{\epsilon}+\frac{3}{4}-\frac{\gamma}{2} \nonumber\\
&-&\frac{1}{2}\text{ln}\bigg(\frac{2eB}{4\pi\mu^2}\bigg)-\frac{1}{2}\text{ln}(x)\bigg]
\end{eqnarray}

Now we add and subtract ~\ref{eq37} from ~\ref{eq35}, we get the regularized pressure with the vacuum part, and the magnetic field-dependent part separated as
\begin{eqnarray}
\label{eq38}
P_{\text{vac}}(S=1/2,B)&=&P_{\text{vac}}(1/2,B=0)+\Delta P_{\text{vac}}(1/2,B) \nonumber\\
\end{eqnarray}
 
where
\begin{eqnarray}
\label{eq39}
\Delta P_{\text{vac}}(S=1/2,B)&=&\frac{(eB)^2}{4\pi^2}\bigg[-\frac{2}{12\epsilon}+\frac{\gamma}{12}\nonumber\\
&+&\frac{1}{12}\text{ln}\bigg(\frac{m^2}{4\pi\mu^2}\bigg)+\frac{x}{2}\text{ln}(x)-\frac{x^2}{2}\text{ln}(x)\nonumber\\
&+&\frac{x^2}{4}-\frac{\text{ln}(x)+1}{12}+\zeta^{'}(-1,x)\bigg]
\end{eqnarray}
 
The field contribution given by ~\ref{eq39} is however divergent due to existence of the magnetic field dependent term $\frac{B^2}{\epsilon}$ ~\cite{Schwinger:1951nm,Elmfors:1993bm,Andersen:2011ip}. We nullify this divergence by redefining field-dependent pressure contribution by incorporating the magnetic field contribution to it 

\begin{equation}
\label{eq40}
\Delta P_{\text{vac}}^{r}=\Delta P_{\text{vac}}(B)-\frac{B^2}{2}
\end{equation}
 
 The divergences are absorbed into the renormalization of the electric charge and the magnetic field strength ~\cite{Endrodi:2013cs},

\begin{equation}
\label{eq41}
B^2=Z_{e}B_r^2; \hspace{0.5cm} e^2=Z_e^{-1}e_r^2; \hspace{0.5cm} e_rB_r=eB
\end{equation}
 
 where the electric charge renormalization constant is
 
\begin{eqnarray}
\label{eq42}
Z_{e}\bigg(S=\frac{1}{2}\bigg)=1+\frac{1}{2}e_r^2\bigg[-\frac{2}{12\epsilon}+\frac{\gamma}{12}+\frac{1}{12}\text{ln}\bigg(\frac{m_{*}}{4\pi\mu^2}\bigg)\bigg],\nonumber\\
\end{eqnarray}
 
where, we fix $m_{*}=m$, {\it i.e} to the physical mass of the particle. Thus the renormalized field dependent pressure without pure magnetic filed ($\frac{B^2}{2}$) contribution is
 
 \begin{eqnarray}
\label{eq43}
\Delta P_{\text{vac}}^r(S=1/2,B)&=&\frac{(eB)^2}{4\pi^2}\bigg[\zeta^{'}(-1,x)+\frac{x}{2}\text{ln}(x)\nonumber\\
&-&\frac{x^2}{2}\text{ln}(x)+\frac{x^2}{4}-\frac{\text{ln}(x)+1}{12}\bigg]\nonumber\\
\end{eqnarray}
 
The renormalized magnetic field ($B$) dependent pressure for spin-zero and spin one can be found using the above similar technique. These terms play an important role in deciding the magnetization of the hadronic matter. We note here that the vacuum pressure depends on the charge, mass, spin, etc. Therefore the total vacuum pressure of hadron gas is estimated by adding the vacuum pressure of all the particles considered in this work.

For spin-zero particle:
\begin{eqnarray}
\label{eq44}
\Delta P_{\text{vac}}^r(s=0,B)&=&-\frac{(eB)^2}{8\pi^2}\bigg[\zeta^{'}(-1,x+1/2)-\frac{x^2}{2}\text{ln}(x)\nonumber\\
&+&\frac{x^2}{4}+\frac{\text{ln}(x)+1}{24}\bigg]
\end{eqnarray}

Similarly, for spin-one particle:
\begin{eqnarray}
\label{eq45}
\Delta P_{\text{vac}}^r(s=1,B)&=& -\frac{3}{8\pi^2}(eB)^2\bigg[\zeta^{'}(-1,x-1/2)\nonumber\\
&+&\frac{(x+1/2)}{3}\text{ln}(x+1/2)\nonumber\\
&+&\frac{2}{3}(x-1/2)\text{ln}(x-1/2)-\frac{x^2}{2}\text{ln}(x)\nonumber\\
&+&\frac{x^2}{4}-\frac{7}{24}(\text{ln}(x)+1)\bigg]
\end{eqnarray}
 
 So,
\begin{eqnarray}
\label{eq46}
\Delta P_{\text{vac}} &=& \Delta P_{\text{vac}}^r(s=0,B)+\Delta P_{\text{vac}}^r(S=1/2,B)\nonumber\\
&+&\Delta P_{\text{vac}}^r(s=1,B)\nonumber\\
\end{eqnarray}

Now after calculating the total vacuum pressure, the vacuum magnetization of the system is given by,
\begin{equation}
\label{eq47}
\Delta M_{\text{vac}} = \frac{\partial (\Delta P_{\text{vac}})}{\partial (eB)} 
\end{equation}
  
With the above-detailed formalism, we now move to discuss the results in the next section. 

\section{Results and Discussion}
\label{res}

\begin{figure*}[ht!]
\begin{center}
\includegraphics[scale = 0.44]{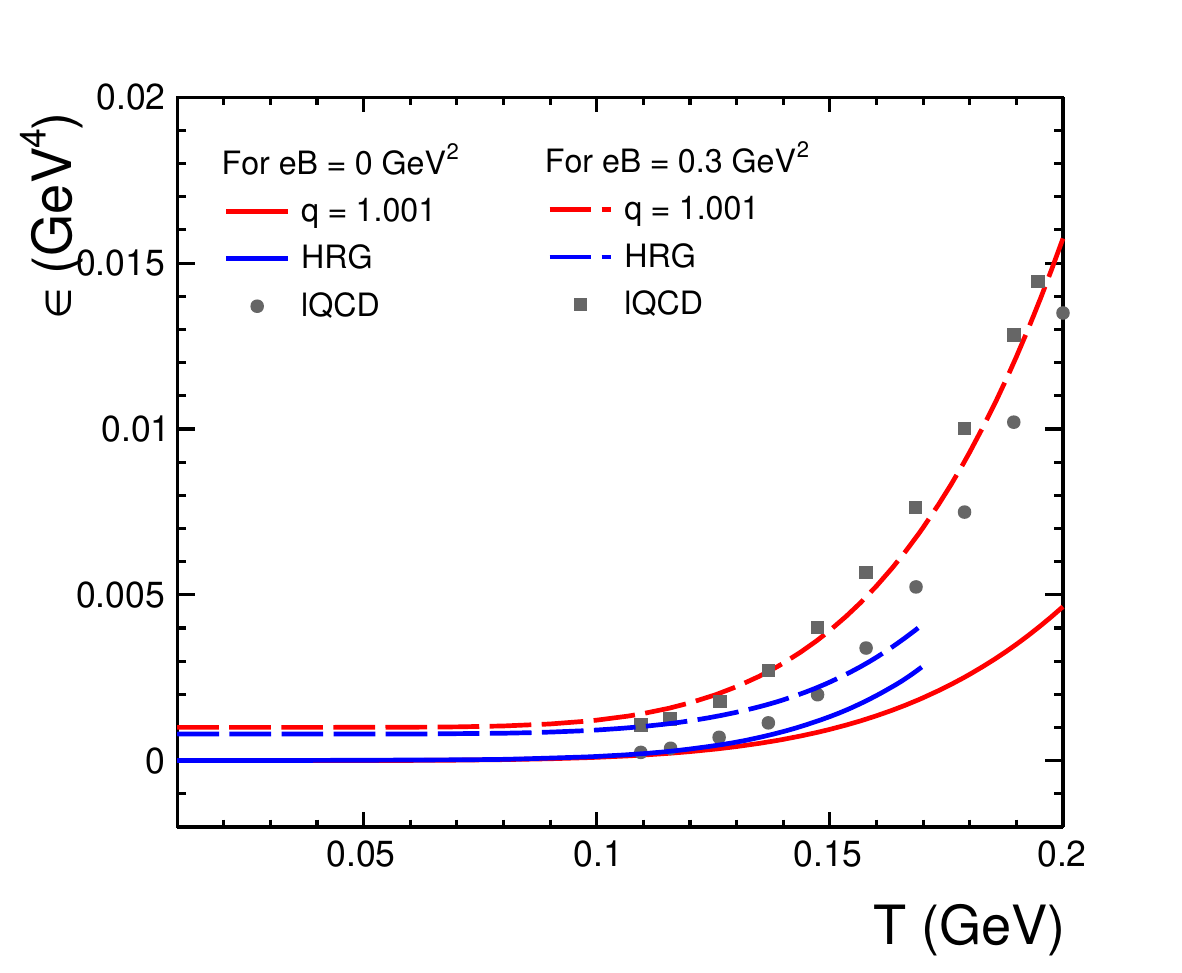}
\includegraphics[scale = 0.44]{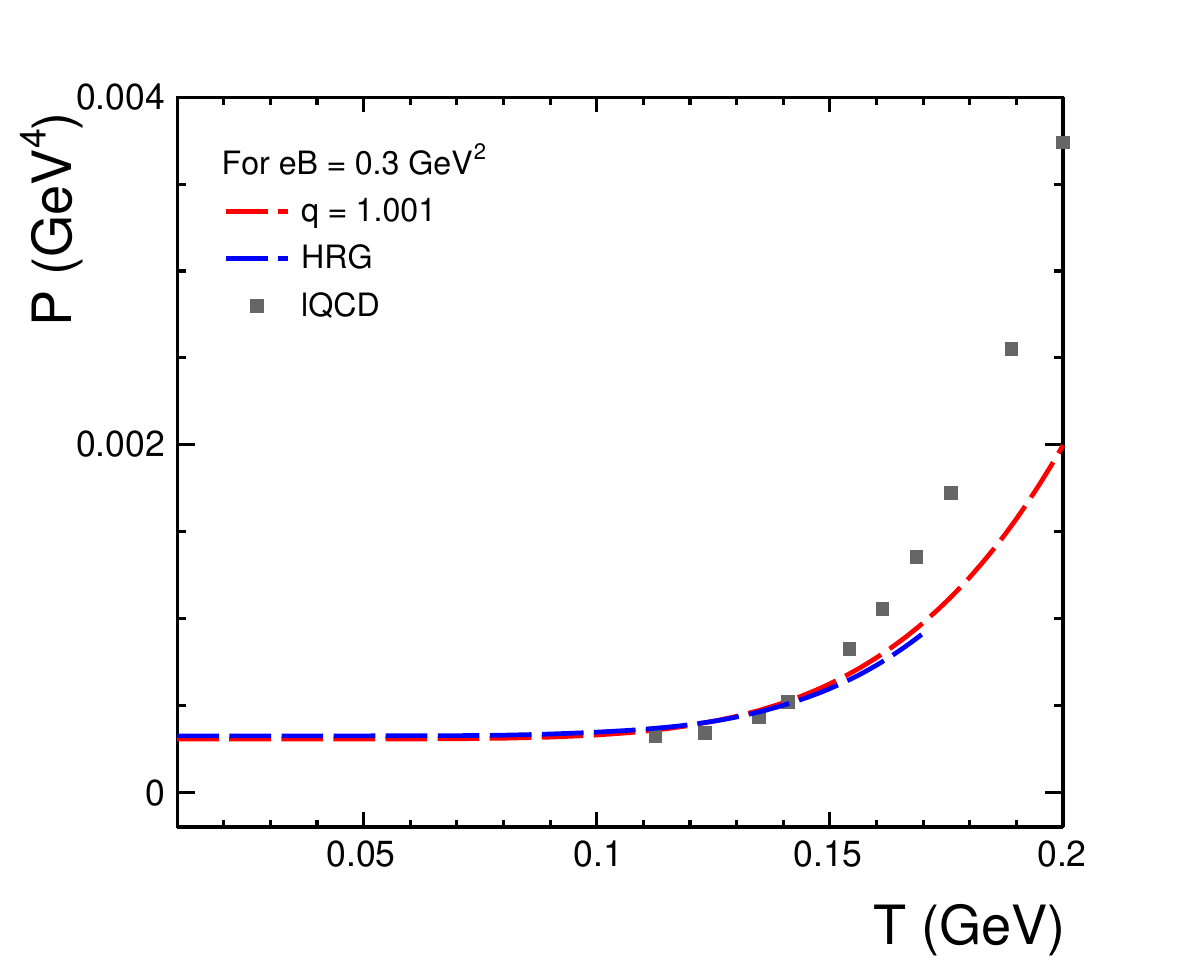}
\caption{(Color online) The left panel shows the total energy density of a hadron gas as a function of temperature for near equilibrium, {\it i.e} for $q$ = 1.001, values of non-extensive parameter at a constant magnetic field $ eB = 0~ {\rm GeV}^{2}$ and $ eB = 0.3 ~{\rm GeV}^{2}$ respectively, compared with corresponding HRG and lattice QCD data ~\cite{Endrodi:2013cs,Bali:2014kia}. The right panel shows the variation of total pressure of a hadron gas as a function of temperature for near equilibrium (for $q$ = 1.001) at a constant magnetic field $ eB = 0.3~ {\rm GeV}^{2}$.}
\label{fig_comparison}
\end{center}
\end{figure*}

We have included the contribution of an external magnetic field in the Tsallis distribution function to calculate the thermodynamical observables of a hadron gas. To validate our calculation for a hadron gas, we have compared our results with the existing results from HRG and lattice QCD model ~\cite{Endrodi:2013cs,Bali:2014kia}. On the left panel of fig.~\ref{fig_comparison}, we have shown the variation of total energy density of a hadron gas as a function of temperature for near equilibrium, {\it i.e} for $q$ = 1.001, values of non-extensive parameter at a constant magnetic field $ eB = 0~ {\rm GeV}^{2}$ and $ eB = 0.3~ {\rm GeV}^{2}$ respectively. We compare our result with the existing HRG and lattice QCD results. For $ eB = 0~ {\rm GeV}^{2}$ our results are in good agreement with the HRG results ~\cite{Endrodi:2013cs} upto $T \sim 150$ MeV. For $ eB = 0.3~ {\rm GeV}^{2}$ the value of total energy density are good in agreement with HRG results upto $T \sim 110$ MeV. However, for an external magnetic field $ eB = 0.3~ {\rm GeV}^{2}$, our results are comparable with the lattice QCD results ~\cite{Bali:2014kia}. Similarly, the right panel of fig.~\ref{fig_comparison} shows the variation of total pressure of a hadron gas as a function of temperature for $q$ = 1.001 at a constant magnetic field $ eB = 0.3~ {\rm GeV}^{2}$. We observe that our calculated total pressure matches with the existing HRG results. However, our estimations and the existing HRG results deviate from the lattice QCD results as we move towards a higher temperature regime.

\begin{figure*}[ht!]
\begin{center}
\includegraphics[scale = 0.44]{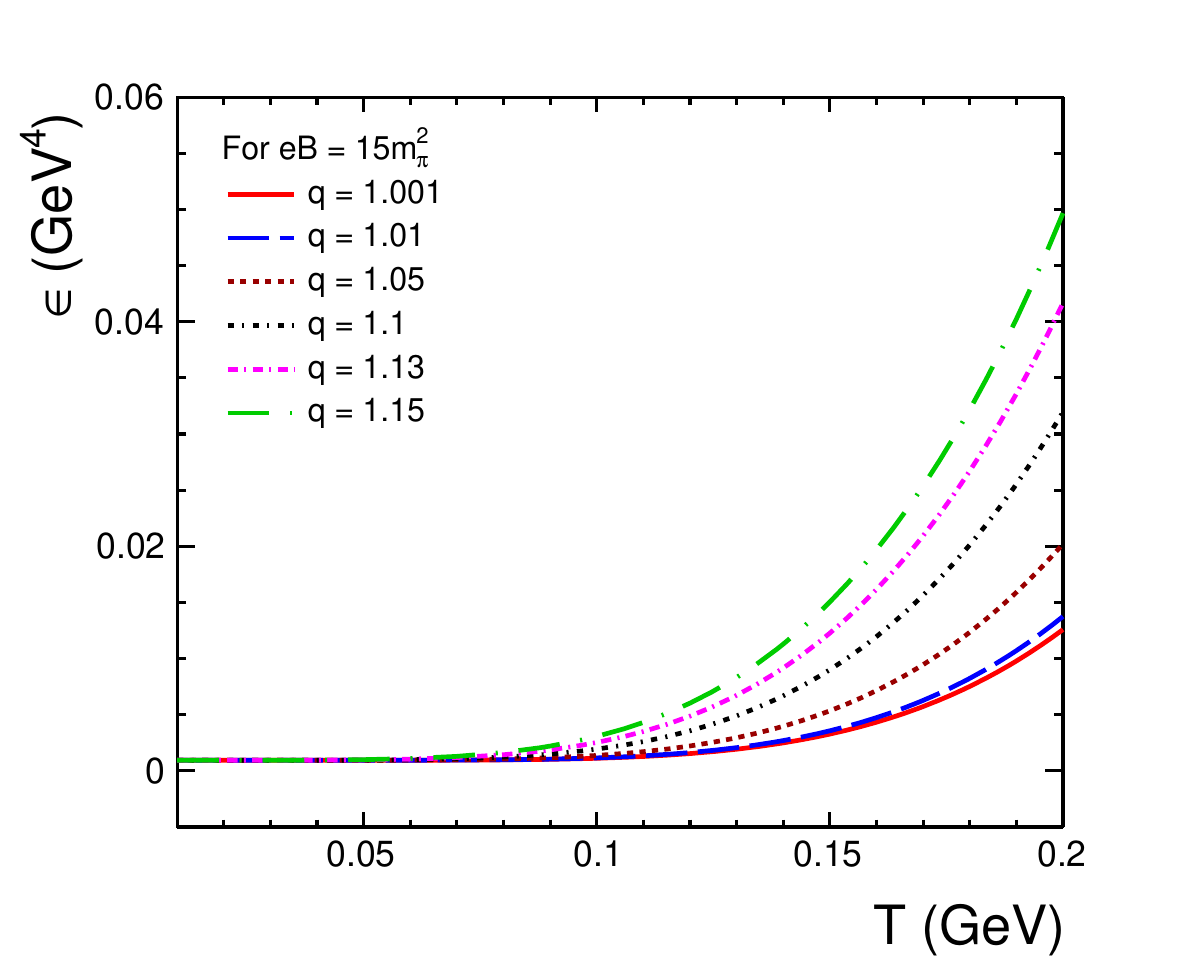}
\includegraphics[scale = 0.44]{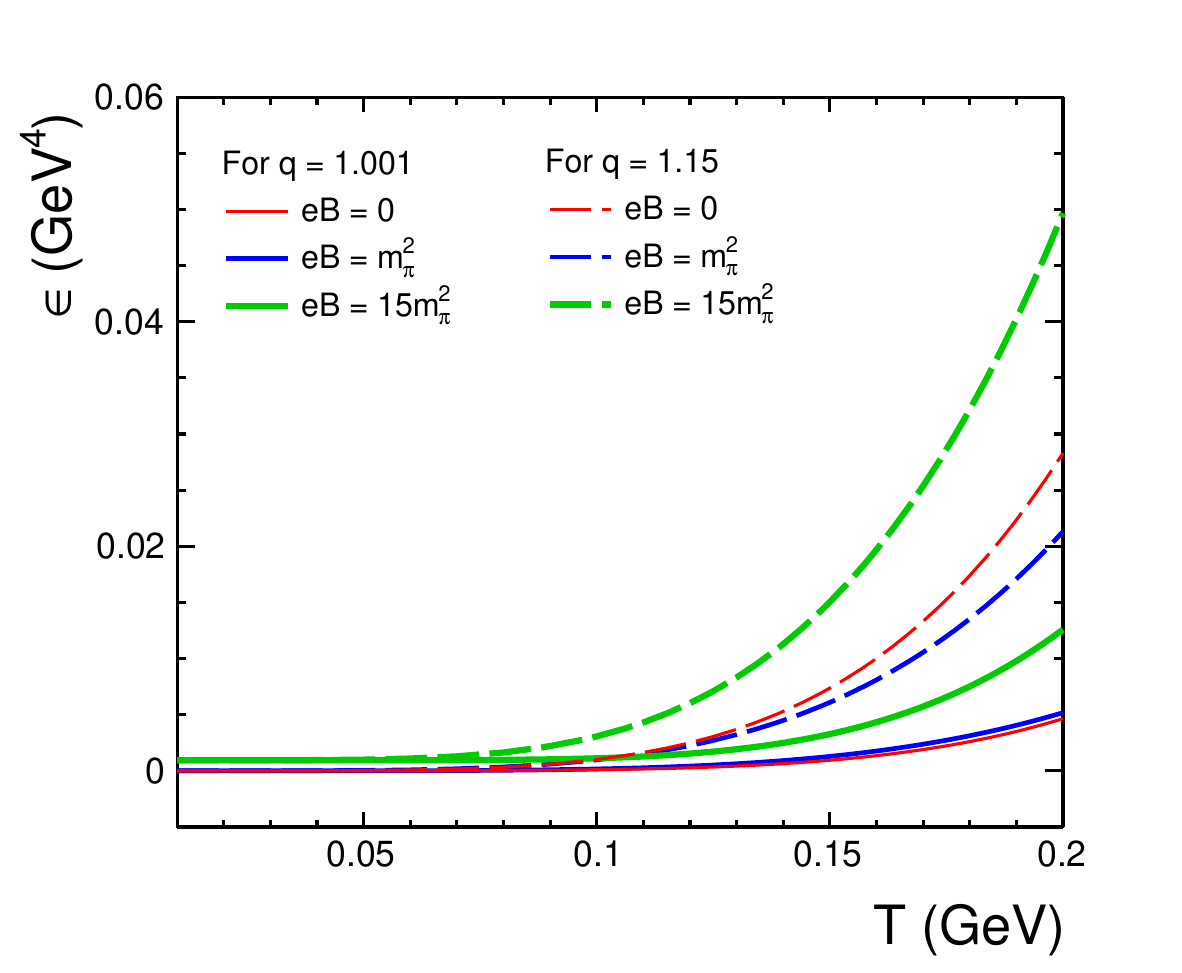}
\caption{(Color online) The left panel shows the variation of total energy density of a hadron gas as a function of temperature for different values of non-extensive parameter at a constant magnetic field $eB = 15m_{\pi}^{2}$. The right panel shows the variation of total energy density at different magnetic fields and different $q$ values.}
\label{fig2}
\end{center}
\end{figure*}

The left panel of fig.~\ref{fig2} shows the variation of the total energy density of a hadron gas as a function of temperature for different values of the non-extensive parameter at a constant external magnetic field of $eB = 15m_{\pi}^{2}$. 
 This value of the magnetic field corresponds to the one usually created in non-central heavy-ion collisions at the LHC
 energies. We find that energy density monotonically increases with an increase in temperature for all $q$-values. At all temperatures, the system with the lowest $q$-value, {\it i.e} the system which is near equilibrium, has the lowest energy density. As the system deviates from the equilibrium, the $q$-value increases, and the energy density of the system also increases. For the system which is far away from equilibrium with a maximum $q$-value, the energy density of the system is the highest at all temperatures. At low temperatures, finite energy density is present due to vacuum contribution.
 
 After observing a strong non-extensive parameter dependence of energy density of the system in the presence of an external magnetic field, now we proceed to take two extreme values of the non-extensive parameter, $q$, to study the competing effects of an external magnetic field and the non-extensive parameter. This is shown in the right panel of fig.~\ref{fig2}. The chosen two extreme values of $q$ are: $q$ = 1.001, which corresponds to a near-equilibrium system and $q$ = 1.15, which corresponds to a system away from equilibrium. In a realistic heavy-ion collision scenario, the former corresponds to central heavy-ion collisions with the creation of an equilibrated system, whereas the later corresponds to mostly non-central heavy-ion or hadronic collisions. The energy density increases with an increase in the magnetic field both for the equilibrated system and the system that is away from equilibrium. We observe an enhanced energy density of the system for the case of the highest magnetic field considered in this study and for the system, which is mostly away from equilibrium. When we consider a system away from equilibrium ($q$ = 1.15), we observe that at a lower temperature regime, the energy density of the system increases with an increase in an external magnetic field. But at a higher temperature regime, the energy density of the system with an external magnetic field $eB = m_{\pi}^{2} $ is lower than that of the system in the absence of an external magnetic field. This observation goes inline with Ref. ~\cite{Kadam:2019rzo}. This result is very interesting considering the fact that this behaviour is observed at high $q$-values, {\it i.e} when the system is away from equilibrium. This is not observed for the case of $q= 1.001$, which is an equilibrated system.
 
\begin{figure*}[ht!]
\begin{center}
\includegraphics[scale = 0.44]{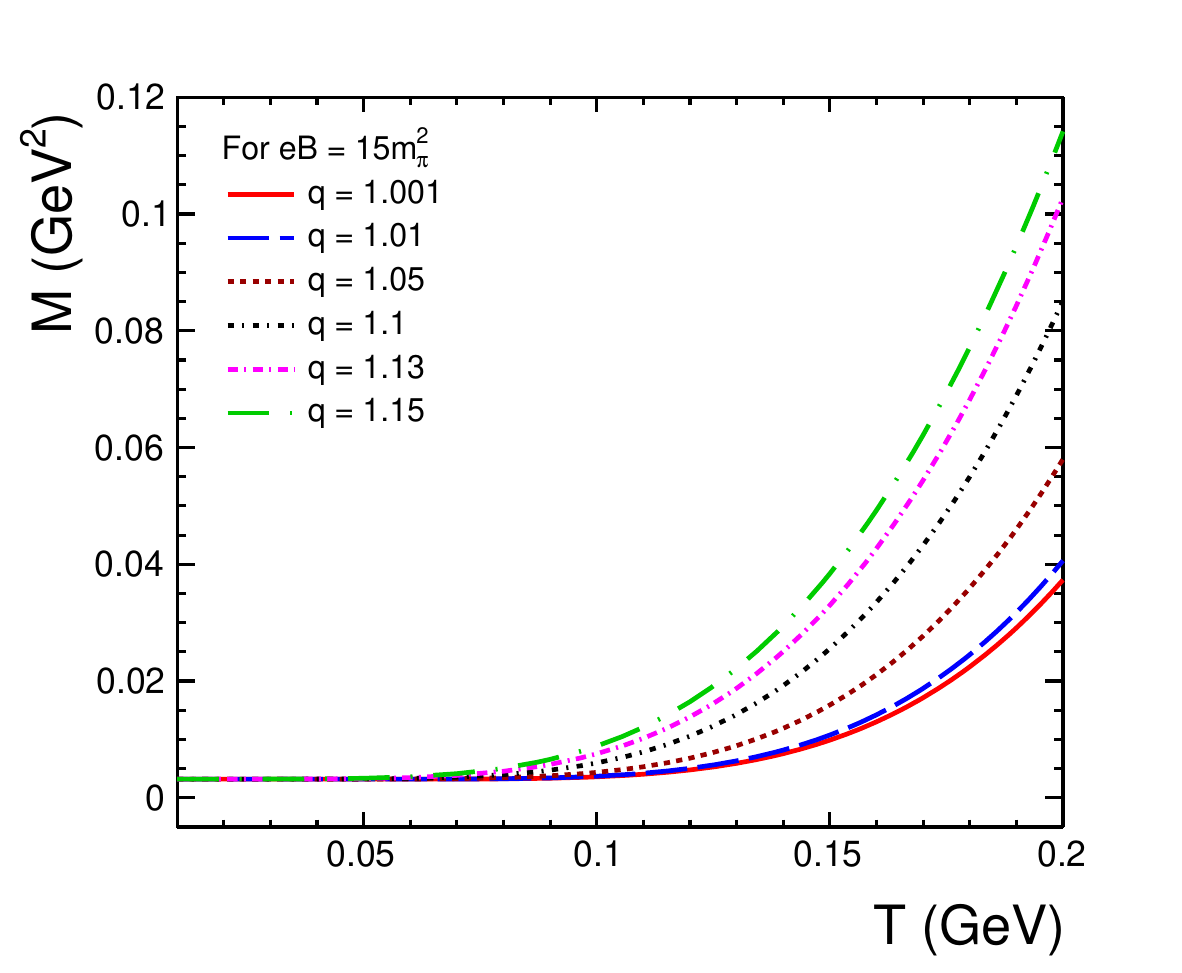}
\includegraphics[scale = 0.44]{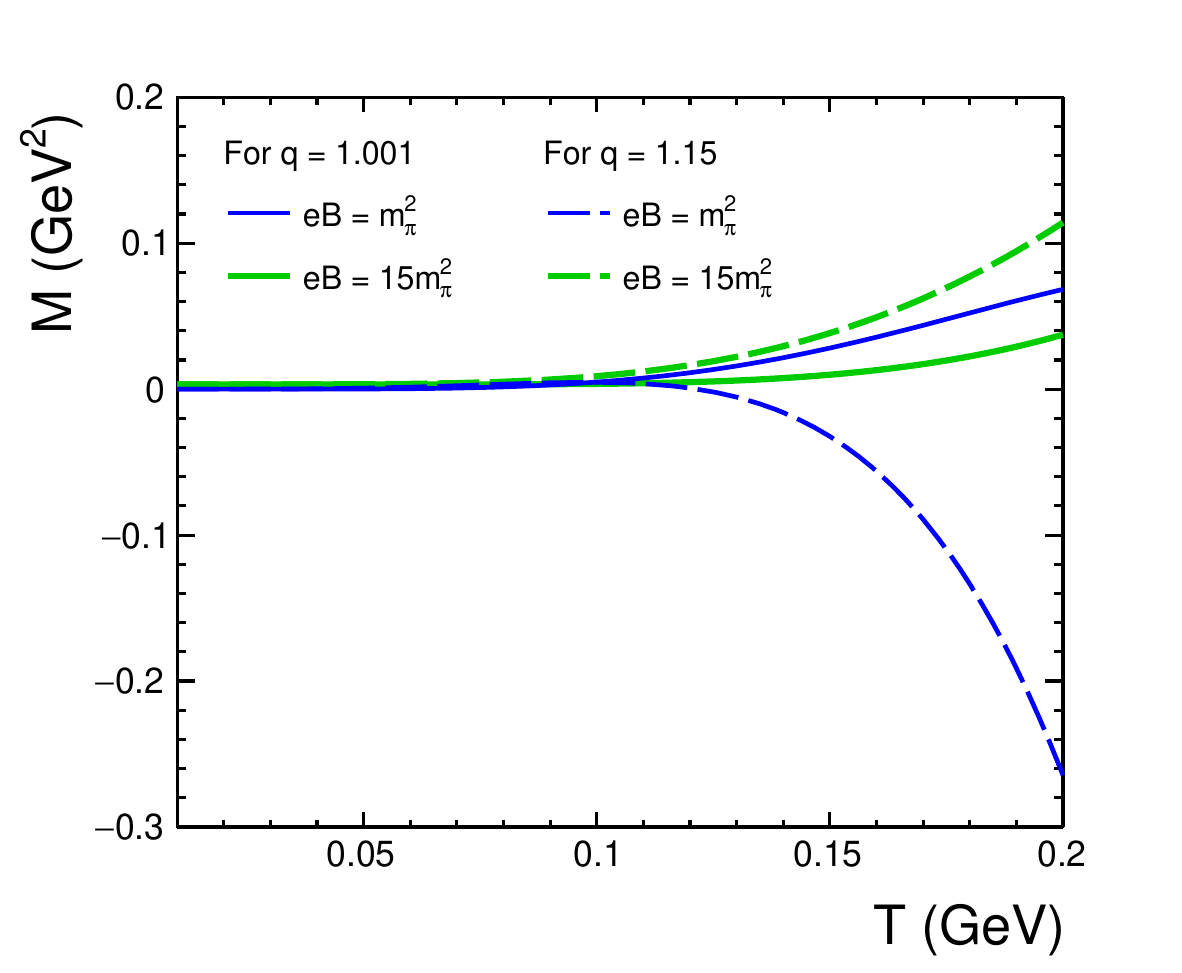}
\caption{(Color online) The left panel shows the variation of magnetization of a hadron gas as a function of temperature for different values of non-extensive parameter at a constant magnetic field $eB = 15m_{\pi}^{2}$. The right panel shows the variation of magnetization at different magnetic fields and for different $q$-values.}
\label{fig3}
\end{center}
\end{figure*}

The left panel of fig.~\ref{fig3} shows the variation of magnetization of a hadron gas as a function of temperature for certain values of the non-extensive parameter at a constant magnetic field $eB = 15m_{\pi}^{2}$. We observe that magnetization increases with an increase in temperature for all $q$-values. At low temperatures, finite magnetization is present due to vacuum contribution and gradually increases with an increase in temperature. For lower $q$-values, {\it i.e} when the system is near equilibrium, the magnetization of the system is lower. As the $q$-value increases, the magnetization of the system increases at all temperatures. For a given temperature of the system, the magnetization is higher for the case of away from equilibrium. 

The right panel of fig.~\ref{fig3} shows the variation of magnetization at different magnetic fields for systems near and away from equilibrium. For $q$ = 1.001 values, when the system is near equilibrium, the magnetization of the system with a lower external magnetic field is higher for all temperatures up to $T \sim 200$ MeV. However, for $q$ = 1.15, when the system is away from equilibrium, the magnetization of the system with external magnetic field $eB = m_{\pi}^{2}$ is observed to be negative. This means that the system produced in high-energy collisions under an external magnetic field has diamagnetic properties. On the other hand, when we increase the external magnetic field for the system away from equilibrium, {\it i.e.} $eB = 15m_{\pi}^{2}$, the system shows positive magnetization. This means that the system has paramagnetic behaviour. This observation is highly interesting as for peripheral heavy-ion collisions, with the increased center-of-mass energy of the system from RHIC to the LHC, the produced magnetic field increases and the system undergoes a diamagnetic to paramagnetic transition. Recall here that, as discussed in Ref. ~\cite{Endrodi:2013cs}, a positive magnetization at a temperature around 150 MeV has been interpreted as due to the formation of a matter which is paramagnetic in nature. Similar observations are also made in Ref. ~\cite{Tawfik:2016lih}. However, the detailed dynamics of such a transition are beyond the scope of the present study and possibly warrant deeper investigations.

\begin{figure*}[ht!]
\begin{center}
\includegraphics[scale = 0.44]{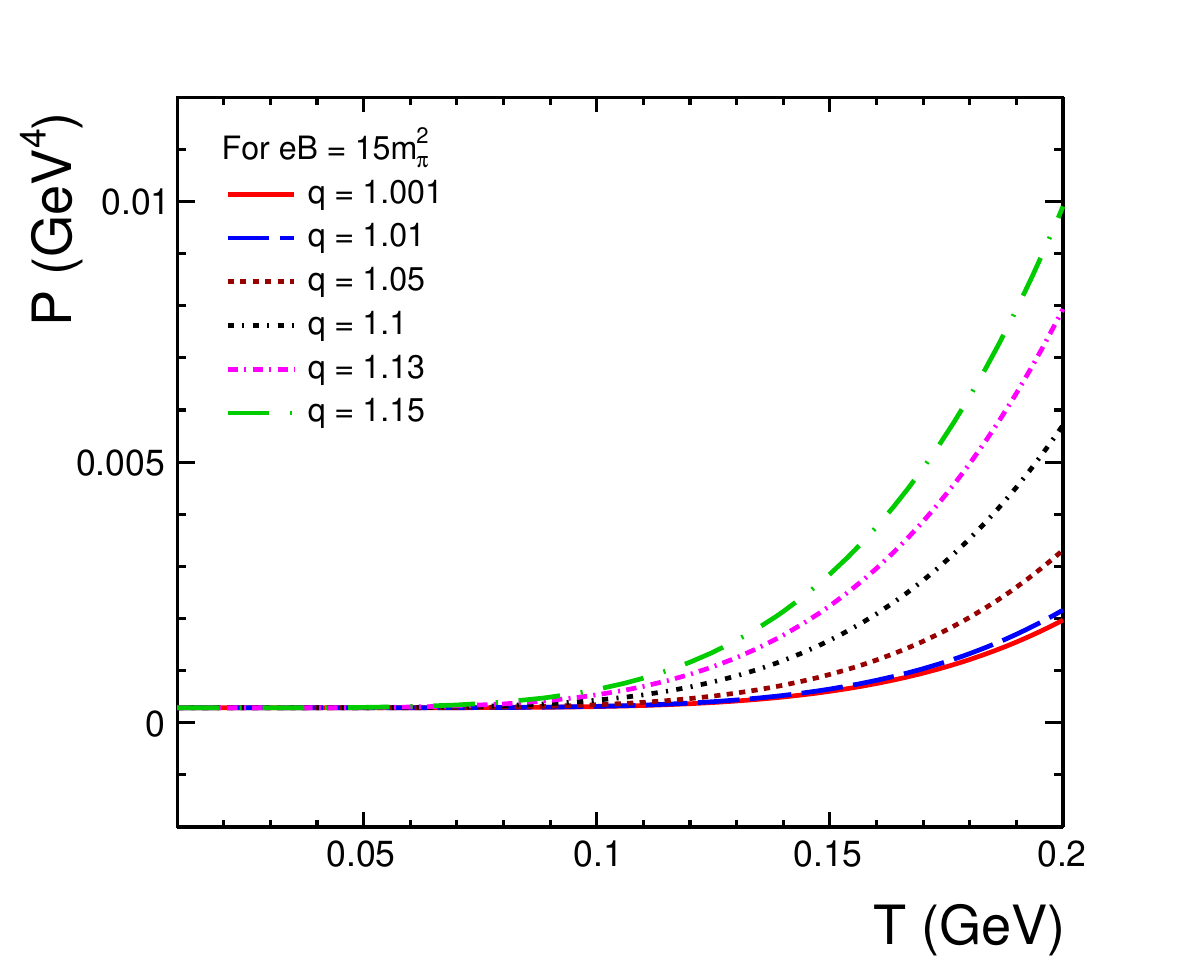}
\includegraphics[scale = 0.44]{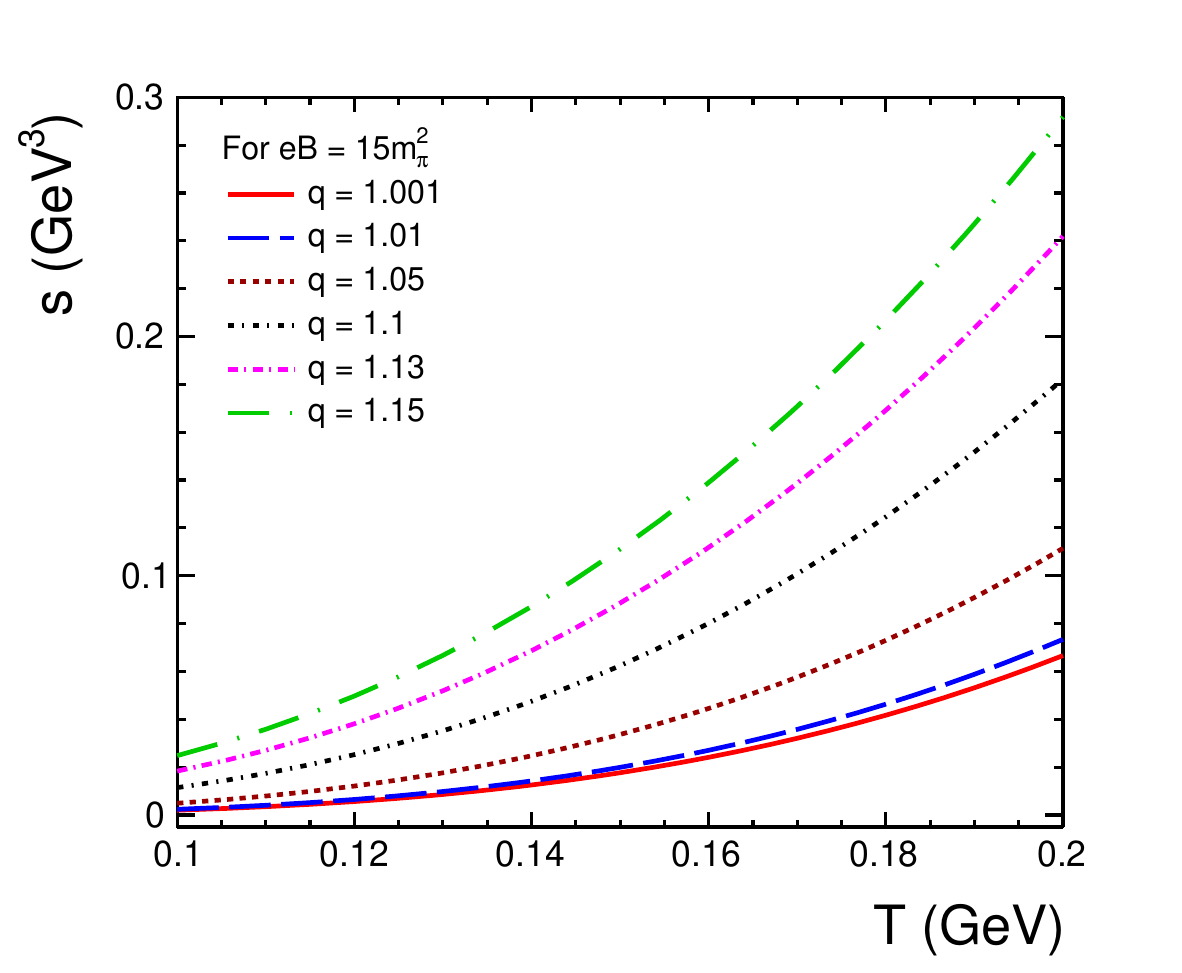}
\caption{(Color online) The left and right panels show the variations of pressure and entropy density as a function of temperature in a magnetized hadron gas for different $q$-values, respectively.}
\label{fig4}
\end{center}
\end{figure*}

The left panel of fig.~\ref{fig4} shows the variation of pressure as a function of temperature in a magnetized hadron gas for different $q$-values. At a finite magnetic field $eB = 15m_{\pi}^{2}$, the pressure of the system increases with an increase in temperature for all the $q$-values. When the system is near equilibrium, {\it i.e} for $q$ = 1.001, the pressure of the system is the lowest and with increasing $q$ values, the pressure of the system increases for all temperatures. This suggests that the non-extensive parameter contributes to the pressure of the system. As we have considered the vacuum contribution in our calculation, the total pressure has a finite value even at zero temperature. The right panel of fig.~\ref{fig4} shows the entropy density of the system as a function of the temperature of a hadron gas under an external magnetic field for different $q$-values. We observe that the entropy density of the system increases with an increase in temperature for all $q$-values. For $q$ = 1.001, when the system is near equilibrium, the entropy density of the system is lowest. As we go away from equilibrium, the entropy density of the system increases for all temperatures, which is intuitive.

\begin{figure}[ht!]
\begin{center}
\includegraphics[scale = 0.47]{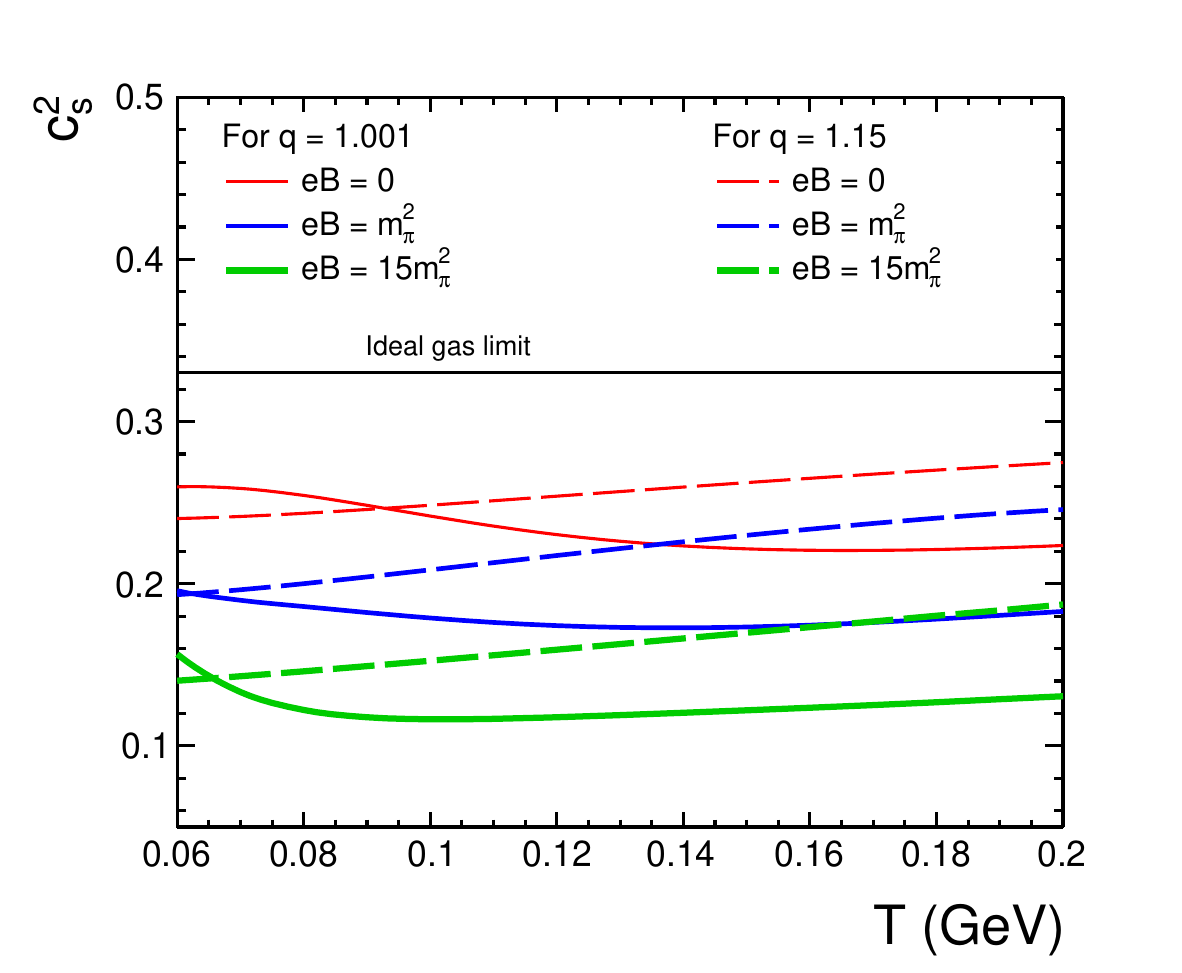}
\caption{(Color online) The squared speed of sound of hadron gas as a function of temperature for different $q$-values for different strengths of magnetic field.}
\label{fig5}
\end{center}
\end{figure}

To understand the hydrodynamical evolution of the matter formed in heavy-ion collisions, one of the most important thermodynamical variables is the speed of sound, $c_{s}$. To explore this, in fig.~\ref{fig5}, we have estimated the squared speed of sound, $(c_{s}^{2})$, for a hadron gas as a function of temperature for different $q$-values under different strengths of external magnetic field ($eB = 0, m_{\pi}^{2}$ and $15m_{\pi}^{2}$). When the system is near equilibrium, {\it i.e} for $q$ = 1.001, the squared speed of sound of the system for all the external magnetic fields is observed to decrease with an increase in temperature. The speed of sound is the highest in the absence of an external magnetic field for the case of a system near equilibrium. When the system is away from equilibrium, {\it i.e} for $q$ = 1.15, we observe an increasing trend with temperature for the squared speed of sound for the system, for almost all the cases of external magnetic fields. The presence of a magnetic field in the system seems to decrease the speed of sound in the medium at all temperatures. This means that under an external magnetic field, the system is more interacting in nature as compared to the system without a magnetic field. External magnetic field sets in the interactions in the system, taking it away from an ideal gas behaviour.

\section{Summary}
\label{sum}
In this work, we have taken the non-extensive Tsallis statistics to study a hadron gas that is formed in peripheral heavy-ion collisions for which the transverse momentum spectra are usually described by non-extensive statistics. Such collisions generate a strong external magnetic field, whose effect can be observed in the final state when we study various thermodynamic properties of the hadron gas. The strength of the produced magnetic field is found to increase with the collision energy of the system, making it the highest one being produced at the LHC. We have studied the effect of the non-extensive parameter, $q$, on various thermodynamic and magnetic properties of the produced system. We observe that when the system is away from equilibrium, it has higher values of energy density, pressure, and entropy density. This suggests that the non-extensive parameter is positively correlated with the above thermodynamic observables. We have also observed that under a very strong magnetic field ($eB = 15m_{\pi}^{2}$), the magnetization of the system under consideration is positive for all values of the non-extensive parameter. However, when the system is away from equilibrium (for $q$ = 1.15), under less strong magnetic field ($eB =m_{\pi}^{2} $), it shows diamagnetic behaviour. This further changes to paramagnetic when the external magnetic field is increased. This indicates a diamagnetic to paramagnetic transition for non-central heavy-ion collisions as one moves from RHIC to the LHC energies. Such an observation is never made so far and possibly needs intense investigations. We have also studied the squared speed of sound of a hadron gas under an external magnetic field. We observe the value of $c_{s}^{2}$ besides respecting the Stefan-Boltzmann limit of 1/3, asymptotically decreases with an increase in magnetic field strength. Hence, the system is more interacting in the presence of a finite magnetic field.

\section*{Acknowledgement} 
This research work has been carried out under financial support from DAE-BRNS, the Government of India,
Project No. 58/14/29/2019-BRNS of Raghunath Sahoo. G.S. Pradhan acknowledges the financial support by the DST-INSPIRE program of the Government of India. R.S. acknowledges the financial support under the CERN Scientific Associateship, CERN, Geneva, Switzerland. The authors further acknowledge Dr. Jayanta Dey and Dr. Arvind Khuntia for their help in carefully reading the manuscript and providing useful comments. The authors gratefully acknowledge the very useful comments from the anonymous referee and the editorial board member.

\section*{Appendix}
\subsection{\label{sec:Consistency}Thermodynamic Consistency of Tsallis distribution function}
 The form of Tsallis distribution function used in the present work is $f^q(E,q,T,\mu)$, which is defined as,

 \begin{equation}
 \label{eq48}
 f^q (E,q,T,\mu) = \frac{1}{[1+(q-1)\frac{E-\mu}{T}]^{\frac{q}{q-1}}}.
 \end{equation} 
 Now we proceed to show its thermodynamic consistency as follows.
 Note that one of the appropriate constraints is given by the average number of particles, {\it i.e.}

 \begin{equation}
 \label{eq49}
\displaystyle\sum_{i} f_i^{q} = N .
 \end{equation} 
 
Correspondingly, the energy of the system gives a constraint, 
\begin{equation}
\label{eq50}
\displaystyle\sum_{i} f_i^{q}E_{i} = E .
\end{equation} 

Here, $E_{i}$ stands for both $E_{i,n}$ and $E_{i,c}(p_{z},k,s_{z})$ and $f_{i}$ stands for $f(E,q,T,\mu)$, which are defined in the formulation part. The first and second laws of thermodynamics follow the following two differential relations ~\cite{deGroot:1980aa}.
 
\begin{equation}
\label{eq51}
d\epsilon = Tds + \mu dn +(eB)dM,
\end{equation} 

\begin{equation}
\label{eq52}
dP = sdT + nd\mu + Md(eB).
\end{equation} 

where $\epsilon = E/V$, $s = S/V$, $n = N/V$ and $M = \mathcal M/V$ are the energy, entropy, particle, and magnetization density respectively.Thermodynamic consistency demands that the subsequent relations be satisfied.

\begin{equation}
\label{eq53}
T = \left.\frac{\partial \epsilon}{\partial s} \right|_{n,M},
\end{equation} 

\begin{equation}
\label{eq54}
n = \left.\frac{\partial P}{\partial \mu}\right|_{T, eB},
\end{equation} 

\begin{equation}
\label{eq55}
s = \left.\frac{\partial P}{\partial T}\right|_{\mu,eB},
\end{equation} 

\begin{equation}
\label{eq56}
 \mu = \left.\frac{\partial \epsilon}{\partial n}\right|_{s,M},
 \end{equation}
 
\begin{equation}
\label{57}
eB = \left.\frac{\partial \epsilon}{\partial M}\right|_{n,s},
\end{equation}
 
\begin{equation}
\label{58}
M = \left.\frac{\partial P}{\partial (eB)}\right|_{T,\mu}.
\end{equation}
 
Now we calculate Eq~\ref{eq53}, which can be explicitly written as,

\begin{eqnarray}
\label{eq59}
 \left.\frac{\partial E}{\partial S} \right|_{n,M} &=& \frac{\frac{\partial E}{\partial T }dT + \frac{\partial E}{\partial \mu }d\mu +\frac{\partial E}{\partial (eB) }d(eB)}{\frac{\partial S}{\partial T}dT + \frac{\partial S}{\partial \mu}d\mu + \frac{\partial S}{\partial (eB) }d (eB)}\nonumber\\
 &=& \frac{\frac{\partial E}{\partial T} + \frac{\partial E}{\partial \mu}\frac{d\mu}{dT}+\frac{\partial E}{\partial (eB)}\frac{d(eB)}{dT}}{\frac{\partial S}{\partial T}+ \frac{\partial S}{\partial \mu}\frac{d\mu}{dT}+\frac{\partial S}{\partial (eB)}\frac{d(eB)}{dT}}
\end{eqnarray}
  
Since $n$ is kept fixed one has the additional constraint, that leads to,
\begin{eqnarray}
\label{eq60} 
dn &=& \frac{\partial n}{\partial T}dT + \frac{\partial n}{\partial \mu}d\mu+\frac{\partial n}{\partial (eB)}d (eB) = 0\nonumber\\
\Rightarrow \frac{d\mu}{dT} &=& -\left[\frac{\frac{\partial n}{\partial T}+\frac{\partial n}{\partial (eB)}\frac{d (eB)}{dT}}{\frac{\partial n}{\partial \mu}}\right].
\end{eqnarray}

Now we calculate each of the terms present in ~\ref{eq59} as follow:

\begin{equation}
\label{eq61}
\frac{\partial E}{\partial T} = \displaystyle\sum_{i} qE_{i}f_i^{q-1}\frac{\partial f_i}{\partial T},
\end{equation}
 
\begin{equation}
\label{eq62}
 \frac{\partial E}{\partial \mu} = \displaystyle\sum_{i} qE_{i}f_i^{q-1}\frac{\partial f_i}{\partial \mu}, 
\end{equation}

\begin{equation}
\label{eq63}
\frac{\partial n}{\partial T} = \frac{1}{V}\left[\displaystyle\sum_{i} qf{_i}^{q-1}\frac{\partial f_i}{\partial T}\right], 
\end{equation}

\begin{equation}
\label{eq64}
\frac{\partial n}{\partial \mu} = \frac{1}{V}\left[\displaystyle\sum_{i} qf{_i}^{q-1}\frac{\partial f_i}{\partial \mu}\right],
\end{equation}

\begin{equation}
\label{eq65}
\frac{\partial S}{\partial T} = \displaystyle\sum_{i} q\left[\frac{-f_i^{q-1}+(1-f_i)^{q-1}}{q-1}\right]\frac{\partial f_i}{\partial T},
\end{equation}

\begin{equation}
\label{eq66}
\frac{\partial S}{\partial \mu} = \displaystyle\sum_{i}q\left[\frac{-f_i^{q-1}+(1-f_i)^{q-1}}{q-1}\right]\frac{\partial f_i}{\partial \mu},
\end{equation}

\begin{equation}
\label{eq67}
\frac{\partial E}{\partial (eB)} = \displaystyle\sum_{i} qE_{i}f_i^{q-1}\frac{\partial f_i}{\partial (eB)},
\end{equation}

\begin{equation}
\label{eq68}
\frac{\partial n}{\partial (eB)} = \frac{1}{V}\left[\displaystyle\sum_{i} qf{_i}^{q-1}\frac{\partial f_i}{\partial (eB)}\right],
\end{equation}

\begin{equation}
\label{eq69}
\frac{\partial S}{\partial (eB)} = \displaystyle\sum_{i} q\left[\frac{-f_i^{q-1}+(1-f_i)^{q-1}}{q-1}\right]\frac{\partial f_i}{\partial (eB)},
\end{equation}

Where,
\begin{equation}
\label{eq70}
\left(1-f_i\right)^{q-1} = f_i^{q-1}\left[1+\frac{(q-1)(E_{i}-\mu)}{T}\right].
\end{equation}
With these substitutions, the numerator of Eq~\ref{eq59} becomes,

\begin{eqnarray}
\label{eq71}
\frac{\partial E}{\partial T} + \frac{\partial E}{\partial \mu}\frac{d\mu}{dT}+ \frac{\partial E}{\partial (eB)}\frac{d (eB)}{dT}\nonumber\\
=\displaystyle\sum_{i} qE_{i}f_{i}^{q-1}\frac{\partial f_{i}}{\partial T} - \displaystyle\sum_{i} qE_{i}f_{i}^{q-1}\frac{\partial f_{i}}{\partial \mu}\left[\frac{\frac{\partial n}{\partial T}+\frac{\partial n}{\partial (eB)}\frac{d (eB)}{dT}}{\frac{\partial n}{\partial \mu}}\right]\nonumber\\
+\displaystyle\sum_{k} qE_{k}f_k^{q-1}\frac{\partial f_k}{\partial (eB)}\frac {d (eB)}{dT}\nonumber\\
=\displaystyle\sum_{i} qE_{i}f_{i}^{q-1}\frac{\partial f_i}{\partial T} - \frac{\displaystyle\sum_{j} qE_{j}f_{j}^{q-1}\frac{\partial f_{j}}{\partial \mu} \displaystyle\sum_{i} qf_{i}^{q-1}\frac{\partial f_{i}}{\partial T}}{\displaystyle\sum_{j} qf_{j}^{q-1}\frac{\partial f_{j}}{\partial \mu}}\nonumber\\
-\frac{\displaystyle\sum_{j} qE_{j}f_{j}^{q-1}\frac{\partial f_{j}}{\partial \mu} \displaystyle\sum_{k} qf_{k}^{q-1}\frac{\partial f_{k}}{\partial(eB)}\frac {d (eB)}{dT}}{\displaystyle\sum_{j} qf_{j}^{q-1}\frac{\partial f_{j}}{\partial \mu}}\nonumber\\
+\displaystyle\sum_{k} qE_{k}f_k^{q-1}\frac{\partial f_k}{\partial (eB)}\frac {d (eB)}{dT}\nonumber\\
= \frac{q\left[\displaystyle\sum_{i,j}E_{i}C_{ij}+\displaystyle\sum_{j,k}E_{j}D_{jk} \frac {d (eB)}{dT}\right]}{\displaystyle\sum_{j} 
f_j^{q-1}\frac{\partial f_j}{\partial \mu}}\nonumber\\
\end{eqnarray}

Here,
\begin{equation}
\label{eq72}
C_{ij}\equiv (f_{i}f_{j})^{q-1}\left[\frac{\partial f_i}{\partial T}\frac{\partial f_j}{\partial \mu} -\frac{\partial f_j}{\partial T}\frac{\partial f_i}{\partial \mu}\right],
\end{equation}

\begin{equation}
\label{eq73}
D_{jk}\equiv (f_{j}f_{k})^{q-1}\left[\frac{\partial f_j}{\partial \mu}\frac{\partial f_k}{\partial (eB)} - \frac{\partial f_k}{\partial \mu}\frac{\partial f_j}{\partial (eB)}\right].
\end{equation}

Similarly, the denominator of Eq~\ref{eq59} is,

\begin{eqnarray}
\label{eq74}
\frac{\partial S}{\partial T} + \frac{\partial S}{\partial \mu}\frac{d\mu}{dT}+\frac{\partial S}{\partial (eB)}\frac{d (eB)}{dT}\nonumber\\
= \frac{\partial S}{\partial T} - \frac{\partial S}{\partial \mu}\left[\frac{\frac{\partial n}{\partial T}+\frac{\partial n}{\partial 
(eB)}\frac{d (eB)}{dT}}{\frac{\partial n}{\partial \mu}}\right]+\frac{\partial S}{\partial (eB)}\frac{d (eB)}{dT}\nonumber\\
= \displaystyle\sum_{i} q\left[\frac{-f_i^{q-1}+(1-f_i)^{q-1}}{q-1}\right]\frac{\partial f_i}{\partial T}\nonumber\\
- \frac{\displaystyle\sum_{j}q\left[\frac{-f_{j}^{q-1}+(1-f_{j})^{q-1}}{q-1}\right]\frac{\partial f_{j}}{\partial \mu}\displaystyle\sum_{i} qf{_i}^{q-1}\frac{\partial f_i}{\partial T}}{\displaystyle\sum_{j} qf{_{j}}^{q-1}\frac{\partial f_{j}}{\partial \mu}}\nonumber\\
-\frac{\displaystyle\sum_{j}q\left[\frac{-f_{j}^{q-1}+(1-f_{j})^{q-1}}{q-1}\right]\frac{\partial f_{j}}{\partial \mu}\displaystyle\sum_{k} qf{_k}^{q-1}\frac{\partial f_k}{\partial(eB)}\frac{d (eB)}{dT}}{\displaystyle\sum_{j} qf{_{j}}^{q-1}\frac{\partial f_{j}}{\partial \mu}}\nonumber\\
+\displaystyle\sum_{k} q\left[\frac{-f_k^{q-1}+(1-f_k)^{q-1}}{q-1}\right]\frac{\partial f_k}{\partial (eB)}\frac{d (eB)}{dT}\nonumber\\
 = \frac{q\left[\displaystyle \sum_{i,j}(E_{i}-\mu)C_{i,j}+\displaystyle \sum_{j,k}(E_{j}-\mu)D_{j,k}\frac{d(eB)}{dT}\right]}{T\displaystyle \sum_{j} f_j^{q-1}\frac{\partial f_j}{\partial \mu}}.\nonumber\\
\end{eqnarray}

Where,
\begin{equation}
\label{eq75}
\frac{-f_i^{q-1}+(1-f_i)^{q-1}}{q-1} = \frac{(E_{i}-\mu)}{T}f_i^{q-1}.
\end{equation}

Now, by substituting Eq~\ref{eq71} and Eq~\ref{eq74} in Eq~\ref{eq59} we obtain,

\begin{equation}
\label{eq76}
\left. \frac{\partial E}{\partial S}\right|_{n,M} = T\left[\frac{\displaystyle\sum_{i,j}E_{i}C_{ij}+\displaystyle\sum_{j,k}E_{j}D_{jk}\frac{d(eB)}{dT}}{\displaystyle\sum_{i,j}(E_{i}-\mu)C_{ij}+\displaystyle\sum_{j,k}(E_{j}-\mu)D_{jk}\frac{d(eB)}{dT}}\right]
\end{equation}

Since,$\displaystyle\sum_{i,j} C_{ij} = 0$ and $\displaystyle\sum_{j,k}D_{jk} = 0$ it eventually leads to the desired result
\begin{equation}
\label{eq77}
\left.\frac{\partial E}{\partial S}\right|_{n,M} = T.
\end{equation}
Another thermodynamic quantity is the number density $n$ and partial derivative with respect to $\mu$ in order to check for thermodynamic consistency. From law of thermodynamics we know that,

\begin{equation}
\label{eq78}
P = \frac{-E_{\text{total}} + TS + \mu N+eB\mathcal M}{V} = \frac{-E + TS + \mu N}{V}, 
\end{equation}

where, $E_{\text{total}} = E + eB\mathcal M$.

\begin{eqnarray}
\label{eq79}
 \left.\frac{\partial P}{\partial \mu}\right|_{T, eB} = \frac{1}{V}\left[-\frac{\partial E}{\partial \mu} +T\frac{\partial S}{\partial \mu} + N+ \mu\frac{\partial N}{\partial \mu}\right]\nonumber\\   
= \frac{1}{V}\left[ N -\frac{\partial E}{\partial \mu} +\frac{\partial E}{\partial \mu}\right].\nonumber\\   
\end{eqnarray}

Where,
\begin{equation}
\label{eq80}
T\frac{\partial S}{\partial \mu} + \mu\frac{\partial N}{\partial \mu}=  \frac{\partial E}{\partial \mu}
\end{equation}

Hence, we obtain,
\begin{equation}
\label{eq81}
\left. \frac{\partial P}{\partial \mu}\right|_{T, eB} = n
\end{equation}

Similarly, we can defined,
\begin{equation}
\label{eq82}
\mathcal M = MV = \displaystyle\sum_{i} f_i^{q}\bigg(\frac{E_{\text{total}}-E_{i}}{eB}\bigg),
\end{equation}

\begin{equation}
\label{eq83}
\frac{\partial E_{\text{total}}}{\partial(eB)} = \displaystyle\sum_{i} qE_{\text{total}}f_i^{q-1}\frac{\partial f_i}{\partial(eB)},
\end{equation}

\begin{equation}
\label{eq84}
\frac{\partial \mathcal M}{\partial(eB)} = \displaystyle\sum_{i} q\bigg(\frac{E_{\text{total}}-E_{i}}{eB}\bigg)f_i^{q-1}\frac{\partial f_i}{\partial(eB)}.
\end{equation}

\begin{eqnarray}
\label{eq85}
 \left.\frac{\partial P}{\partial (eB)}\right|_{T, \mu}\nonumber\\ 
 = \frac{1}{V}\left[-\frac{\partial E_{\text{total}}}{\partial(eB)} +T\frac{\partial S}{\partial(eB)}
 +\mu\frac{\partial N}{\partial(eB)}+\mathcal M+eB\frac{\partial \mathcal M}{\partial(eB)}\right]\nonumber\\ 
= \frac{1}{V}\left[ \mathcal M -\frac{\partial E_{\text{total}}}{\partial(eB)} +\frac{\partial E_{\text{total}}}{\partial(eB)}\right]. \nonumber\\ 
\end{eqnarray}

Where,
\begin{equation}
\label{eq86}
T\frac{\partial S}{\partial(eB)} + \mu\frac{\partial N}{\partial(eB)}+eB \frac{\partial \mathcal M}{\partial(eB)} =  \frac{\partial E_{\text{total}}}{\partial(eB)}
\end{equation}

Hence, we obtain desire result,
\begin{equation}
\label{eq87}
 \left.\frac{\partial P}{\partial (eB)}\right|_{T, \mu} = M
\end{equation}

Hence, it is shown that the definitions of temperature, pressure and magnetization density within the form of Tsallis non-extensive statistics used in this work, are consistent with the first and second laws of thermodynamics. The remaining relations can also be verified in a similar fashion.

{}

\end{document}